\documentclass{article}
\usepackage{graphicx} %
\usepackage{hyperref} %
\usepackage{authblk}
\usepackage{babel}

\title{PWFA linear collider improvements \\ from previous concepts to HALHF} 
\author[1,2]{{Erik Adli}\thanks{Erik.Adli@fys.uio.no}}

\author[1,2]{Carl A. Lindstr{\o}m}
\affil[1]{Department of Physics, University of Oslo, Oslo, Norway}
\affil[2]{on behalf of the HALHF Collaboration}
\date{September 9, 2025}

\begin{document}

\maketitle

\begin{abstract}
    This note summarizes the major design
changes from the plasma wakefield linear collider concept presented at Snowmass 2013 to the most recent HALHF 2.0 baseline. 
\end{abstract}

\section{Introduction}

In this note we address a question concerning plasma wakefield linear colliders (PWFA-LC) that came up during the EPPSU 2025--2026: what are the major design changes from the PWFA-LC concept presented at Snowmass 2013~\cite{PWFALC2013} -- here referred to as simply ``2013'' -- to the most recent HALHF 2.0 baseline~\cite{Foster2025}, submitted to the EPPSU 2025~\cite{HALHF_EPPSU2025, HALHF_EPPSU2025_BU} -- referred to as ``HALHF''. 
\footnote{PWFA experiments have also come much closer to demonstrating collider parameters in the period from~\cite{PWFALC2013} to~\cite{Foster2025}. This is not discussed in this note, which focuses on the collider design aspects.} 

Both authors have been involved in the discussions of a PWFA-LC since 2013. HALHF~\cite{Foster2025} should be seen as a qualitative improvement with respect to 2013~\cite{PWFALC2013}; suggestions for improvement of the latter concept have been taken into account in HALHF, to the extent possible with the manpower available.  Many improvements have been documented in the period between 2013 and 2025~\cite{PWFA_SUMMARY_RS}. However HALHF is the first attempt to integrate these improvements into a self-consistent design. The full documentation of HALHF, including the EPPSU input, can be found in Refs.~\cite{Foster2025, HALHF_EPPSU2025, HALHF_EPPSU2025_BU}. A summary of the main improvements from 2013 to HALHF includes that
\begin{itemize}
    \item positron acceleration in plasma is no longer assumed (an RF linac is used)
    \item the transverse beam break-up instability (BBU) has been taken into account in the design process and mitigated (among other effects, ion motion is used for mitigation)
    \item an emittance-preserving interstage has been designed, using a new achromatic staging lattice
    \item a lower drive-beam energy, and a CLIC-like drive-beam concept, with a significantly reduced driver spacing (using CLIC-like combiner rings), have been adopted
    \item plasma relaxation and cooling have been taken into account in the parameter optimization and choice of time-structure
    \item new tools have made integrated simulations possible
    \item a global system optimization for cost has been applied to the system design
    \item a first cost estimate, building on the CLIC/ILC costing, has been performed
\end{itemize}
The following sections detail each point.

\section{Positron acceleration}

As positron acceleration of collider-quality beams is considered very challenging~\cite{Cao2024}, the major design change made for HALHF is to only accelerate the $e^-$ beam in plasma, while assuming an RF linac for the $e^+$ beam~\cite{HALHF}. The colliding beam energies are asymmetric in order to reduce the overall footprint and cost. The HALHF study indicates that such asymmetric collisions should still give good physics performance~\cite{HALHF}. The combination of an $e^-$ plasma linac and $e^+$ RF linac is not trivial, and constrains the overall parameters. The HALHF optimised design approach (see below) takes these constraints into account. 

\section{Transverse BBU instability}

The 2013 parameters~\cite{PWFALC2013}, while resulting in 50\% drive-to-main-beam efficiency, leads to an excessive BBU-rate. Taking into account the efficiency--instability relation~\cite{Finnerud2025, Diederichs2024}, and reducing the charge, the bunch length, and also the efficiency (now at 40\%), while simultaneously adding a controlled amount of ion motion for further mitigation, the BBU-rate in HALHF is significantly lower.  Work is under way to quantify the resulting tolerances~\cite{AdliIPAC2025}.

\section{Interstage design}

In 2013, an interstage design was not completed.  For HALHF, a full interstage has been designed, and implemented in the integrated simulations discussed below. The interstage design preserves transverse emittances, using novel achromatic lattice design based on nonlinear optics, and performs a longitudinal self-correction leading to low energy spread and significantly improved synchronization tolerances~\cite{CARL_selfcorrection, Drobniak2025}. Synchrotron radiation effects are mitigated (verified in simulations) by suitably ramping down the magnetic fields of the interstage dipoles.  The interstage design is currently being further refined under an ERC-grant~\cite{SPARTA}.

\section{Drive-beam}

In 2013 the drive-beam energy was 25 GeV.  This caused too much synchrotron radiation in the accumulator ring and in the delay chicanes.  In HALHF the drive-beam energy is 4 GeV (as compared to 2.4 GeV in CLIC), mitigating sufficiently the synchrotron-radiation issues.  The cost is an increased number of stages, which due to interstage optics reduces the effective acceleration gradient, but this was accounted for as part of the cost optimization routine.  

In 2013, a SCRF drive-beam linac was assumed, without being designed in any detail.  The drive-beam in HALHF is generated by a drive-beam scheme very similar to CLIC, building upon the design-work done by CLIC, while carefully optimizing the parameters to give a time-structure that minimizes the HALHF cost~\cite{Foster2025,HALHF_EPPSU2025, HALHF_EPPSU2025_BU}. The main difference is the use of 4 ns spacing  compared to CLIC's 1 ns (allowing a higher beam-loaded RF gradient, while only reducing the wall-plug-to-driver efficiency by around 5\%). The use of combiner rings (HALHF adopts the same compression factor as CLIC) reduces the driver separation by a factor 24 (from 4 ns to 0.167 ns) -- this significantly reduces the size of the driver delay chicanes. A consequence is that the drive beam injection must be based on RF deflectors instead of fast kickers.

A SCRF drive-beam linac could also be an option for HALHF~\cite{Foster2025}, but presently the warm RF design results in the lowest overall cost.

\section{Plasma relaxation and cooling}

The challenges of plasma-stage cooling, and plasma-relaxation requirements between the bunches, were pointed out in 2013, and have been a key consideration for the HALHF parameter choices.  In particular, high-repetition studies of plasma acceleration is underway~\cite{Darcy2019}, the HALHF time structure is compatible with plasma-relaxation measurements~\cite{Darcy2022} (if assuming light gases such as hydrogen, helium or lithium), and design and tests of cooled plasma cells to mitigate plasma heating are underway at Oxford and DESY. 

\section{Simulation tools}

While PIC simulations of the plasma stage were performed in 2013, simulation tools for integrated simulations of the full plasma linac were not yet developed. For HALHF, the newly developed system code ABEL~\cite{CHEN_ABEL} makes it possible to do integrated simulations of a plasma linac, including the plasma stages, ramps and interstage lattices. PIC codes like HiPACE++~\cite{DID_HIPACE} have also been significantly upgraded in terms of speed (using GPU acceleration) and precision (using mesh refinement), making simulations of beams with HALHF parameters, including crucial effects like ion motions, possible over the full length of a plasma stage.  Integrated start-to-end simulations of the HALHF plasma linac, using ABEL combined with HiPACE++ have been performed \cite{Foster2025,HALHF_EPPSU2025, HALHF_EPPSU2025_BU}.

\section{Optimised system design}

The HALHF parameters\cite{Foster2025,HALHF_EPPSU2025, HALHF_EPPSU2025_BU} are the results of a Bayesian global optimization~\cite{LINDSTROM_IPAC} of the full programme cost, performed with ABEL.  12 parameters were simultaneously varied using the Bayesian optimization:

(1) the energy asymmetry; 
(2) the number of bunches in a train; 
(3) the repetition rate of the trains; 
(4) the combiner-ring compression factor; 
(5) the drive-bunch temporal separation; 
(6) the number of RF cells per structure in the driver linac;
(7) the number of structures per klystron in the driver linac; 
(8) the accelerating gradient in the driver linac; 
(9) the accelerating gradient in the positron linac; 
(10) the number of RF cells per klystron in the positron linac; 
(11) the number of PWFA stages;
(12) the PWFA transformer ratio.

The resulting optimum was consistently found across multiple optimization runs, indicating that it is the global optimum (given these parameters).

\section{Cost estimate}

A first cost estimate has been performed for a CERN-sited HALHF, building on the detailed cost estimates available for CLIC and an SCRF-LC@CERN for the EPPSU 2025~\cite{Foster2025,HALHF_EPPSU2025, HALHF_EPPSU2025_BU}. 

\newpage

\end{document}